\documentclass{emulateapj}
\usepackage{apjfonts}

\shorttitle{The solar hydrogen spectrum in non-LTE}
\shortauthors{Przybilla \& Butler}

\begin{document}

\title{The solar hydrogen spectrum in non-LTE}

\author{Norbert Przybilla}
\affil{Dr. Remeis-Sternwarte Bamberg, Sternwartstrasse 7, D-96049 Bamberg, Germany}
\email{przybilla@sternwarte.uni-erlangen.de}

\and

\author{Keith Butler}
\affil{Universit\"atssternwarte M\"unchen, Scheinerstrasse 1, D-81679 M\"unchen, Germany}
\email{butler@usm.uni-muenchen.de}

\begin{abstract}
We investigate the synthesis of the Balmer and Paschen lines
of the quiet Sun, using both classical semi-empirical and
theoretical model atmospheres, modern line broadening theory and non-LTE 
line-formation. The computations alleviate long-standing discrepancies 
between LTE predictions
and the observed lines. Theoretical and semi-empirical
model atmospheres
without a chromosphere on the one hand and semi-empirical models with a chromosphere
on the other produce two physically disjoint solutions for the run of
non-LTE level populations, including H\,{\sc ii}, throughout the model 
stratification.
The resulting synthetic non-LTE line profiles are practically identical and 
reproduce the
observation in either case, despite large differences in the line-formation
depths, e.g., a chromospheric origin of the H$\alpha$ core (in concordance 
with observation) versus a photospheric origin.
The findings are of much broader interest, assuming the Sun to
be a prototype cool dwarf star. A consistent account for chromospheres in 
cool
star analyses is required, due to their potential to change
atmospheric structure via non-LTE effects on the ionization balance of hydrogen
and thus the free electron pool. The latter in turn affects the main
opacity source H$^-$. This will in
particular affect the atmospheres of metal-poor and evolved stars, in which
the contribution of hydrogen to the electron pool becomes dominant.
\end{abstract}


\section{Introduction}
Understanding the Sun is of fundamental importance to stellar astronomy and
astrophysics. 
Naturally, the hydrogen
lines are an important aspect, in particular the first members of the
Balmer line series.
The Balmer lines are vital for tests of the internal consistency of model
atmospheres as they sample the physical conditions throughout the 
stellar atmosphere.
Modelling of the hydrogen line wings allows for an accurate temperature 
determination in cool stars, superior to that achievable using broad- or 
intermediate-band photometric indicators (Fuhrmann, Axer \& Gehren~1993, 1994). 

Traditionally, either
semi-empirical atmospheres (e.g. Maltby et al.~1986, MACKKL; 
Hol\-weger \& M\"uller~1974, HM) -- with or without chromosphere -- 
or models computed from basic physical principles
including local thermodynamic equilibrium and energy flux conservation 
(e.g. Kurucz~1994, SUNK94) are used for analyses. 
In combination with the most recent data on line-broadening, in
particular the Stark broadening tables of Stehl\'e \& Hutcheon~(1999,
SH) and the self-broadening formalism of Barklem, Piskunov \& O'Mara~(2000, BPO),
good fits to the wings of the normalised Balmer lines are obtained (BPO; 
Cowley \& Castelli~2002). However, the models fail to reproduce the
cores of the lines, most notably the Doppler core of H$\alpha$, which are
of chromospheric origin, and the far line wing--continuum
transition of H$\beta$ and the higher Balmer lines, see e.g. BPO. 

Recently, the first sophisticated time-dependent {\em ab-initio} 3D
radiative-hydrodynamical models of the solar atmosphere became available
(Asplund et al.~2000), together with its temporally and spatially
averaged 1D representation (Asplund et al.~2004, A-1D). 
It has been suggested that these solve many
long-standing issues of the quantitative interpretation of the solar
spectrum, and of cool stars (Asplund et al.~2004, and
references therein). On the other hand, in a
more conventional approach, Grupp (2004, MAFAGS-OS) shows that  important improvements can be made when
handling atomic data more realistically, reducing the {\em missing
opacity} problem in the Sun and thus improving the modelling of the solar
flux distribution.

Here, we investigate the relevance of non-LTE computations for the
interpretation of the solar Balmer and Paschen lines as a test case for
solar-type stars in general. The study is motivated by our
findings on the impact of improved collision data on the modelling of the
hydrogen lines in early-type stars (Przybilla \& Butler~2004, PB), but
it will be shown to be more far-reaching than that.
In the following sections we provide details of our model calculations, 
compare with observation and discuss the implications.


\section{Model calculations}
   \begin{figure*}
\plottwo{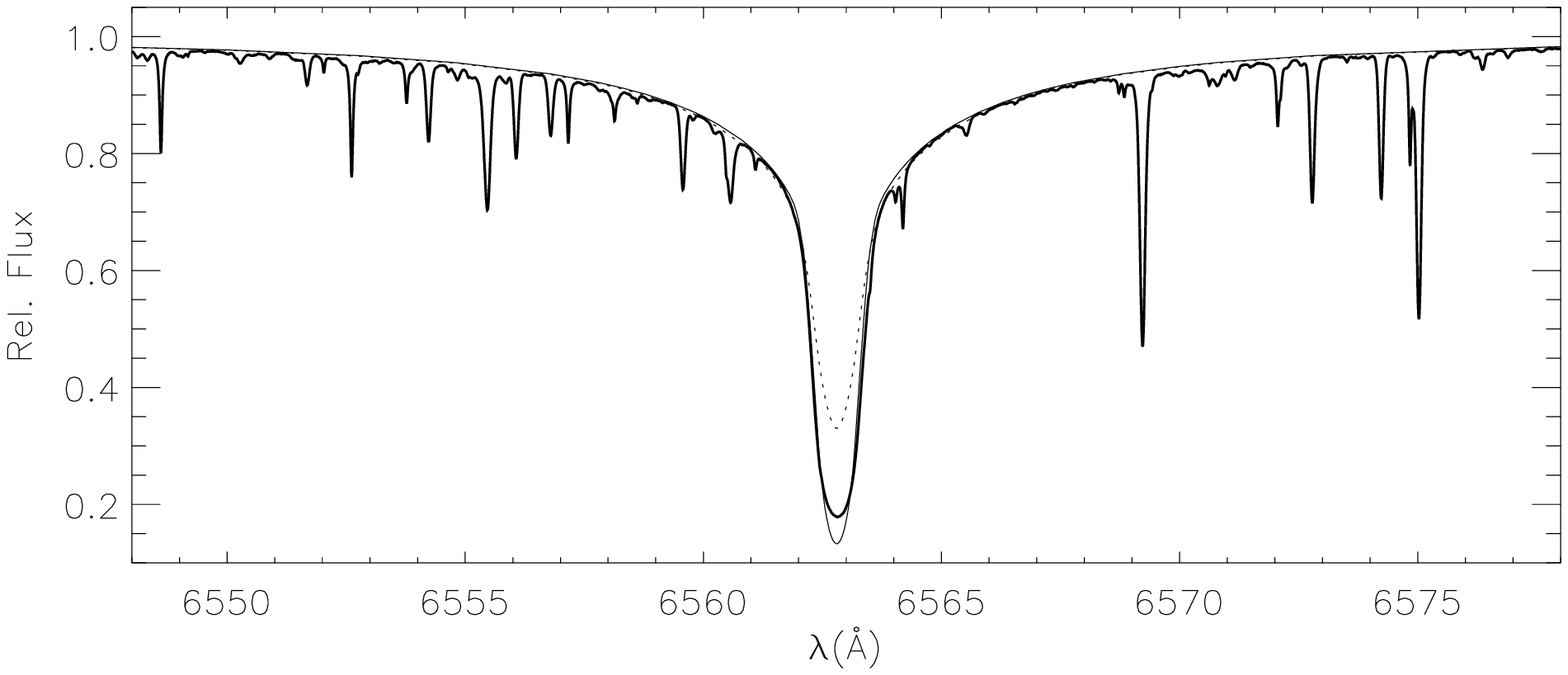}{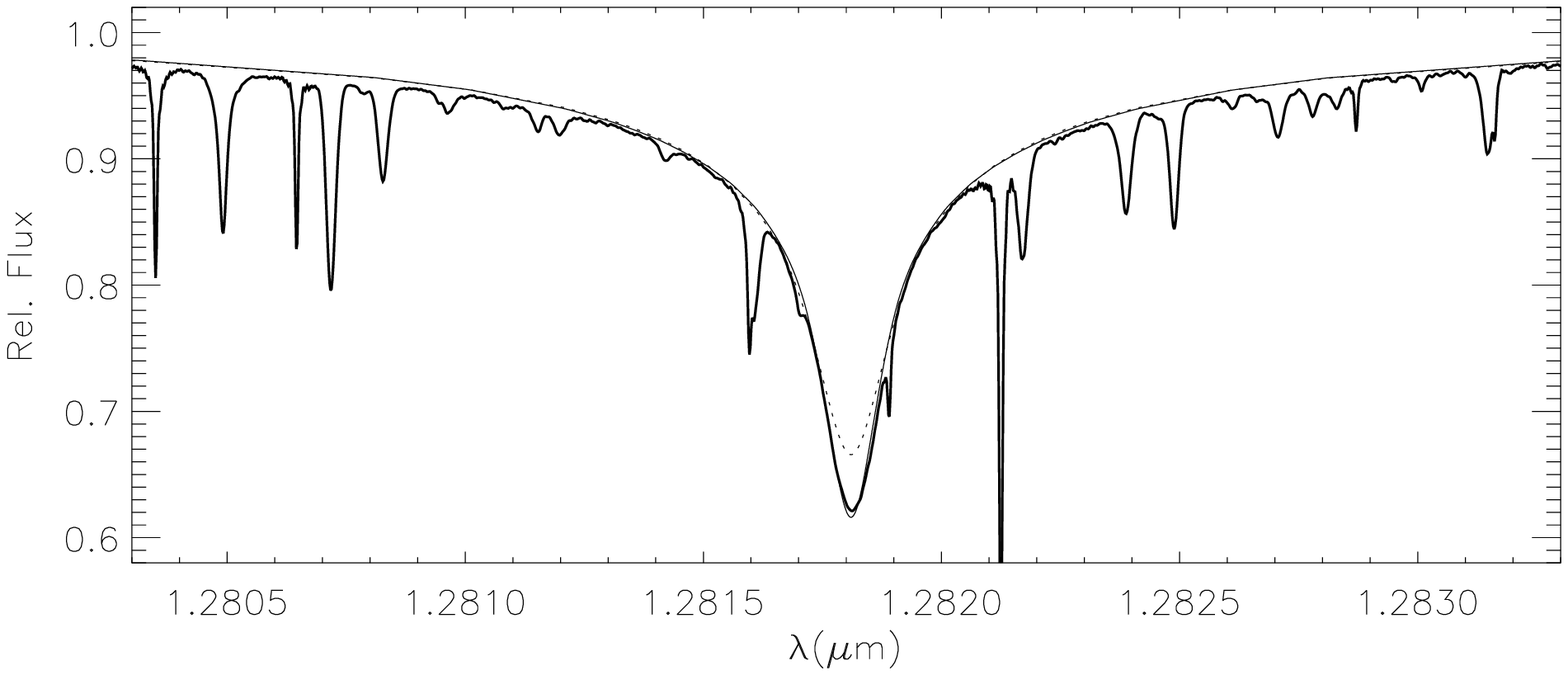}\\
\plottwo{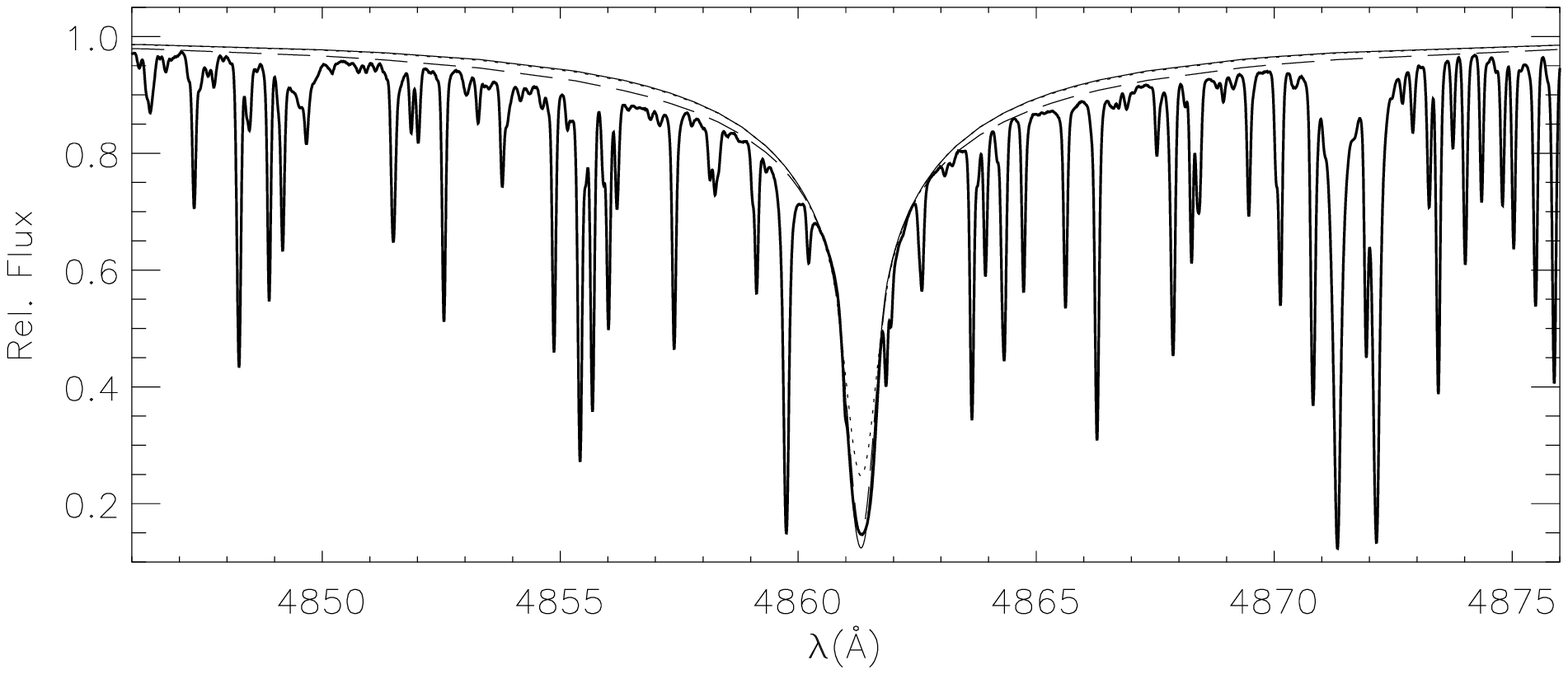}{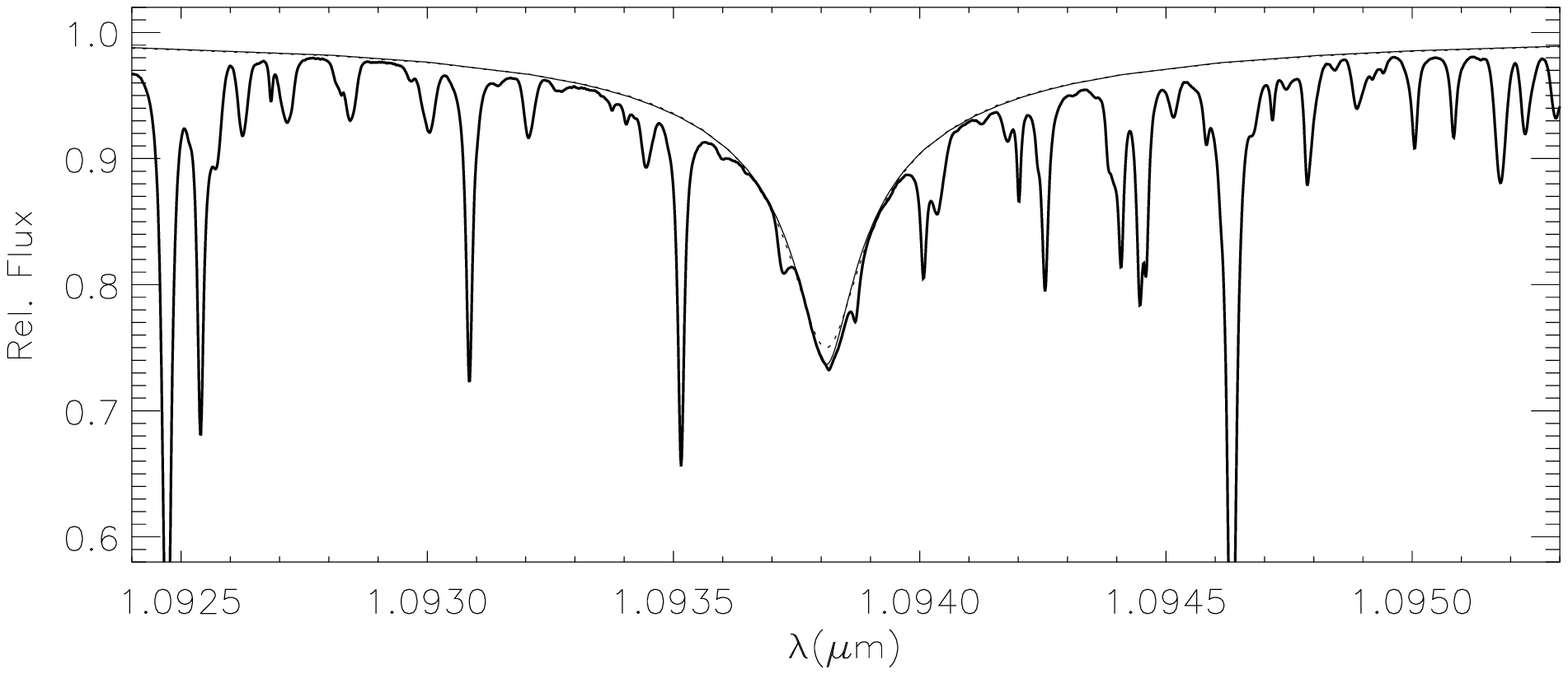}\\
\plottwo{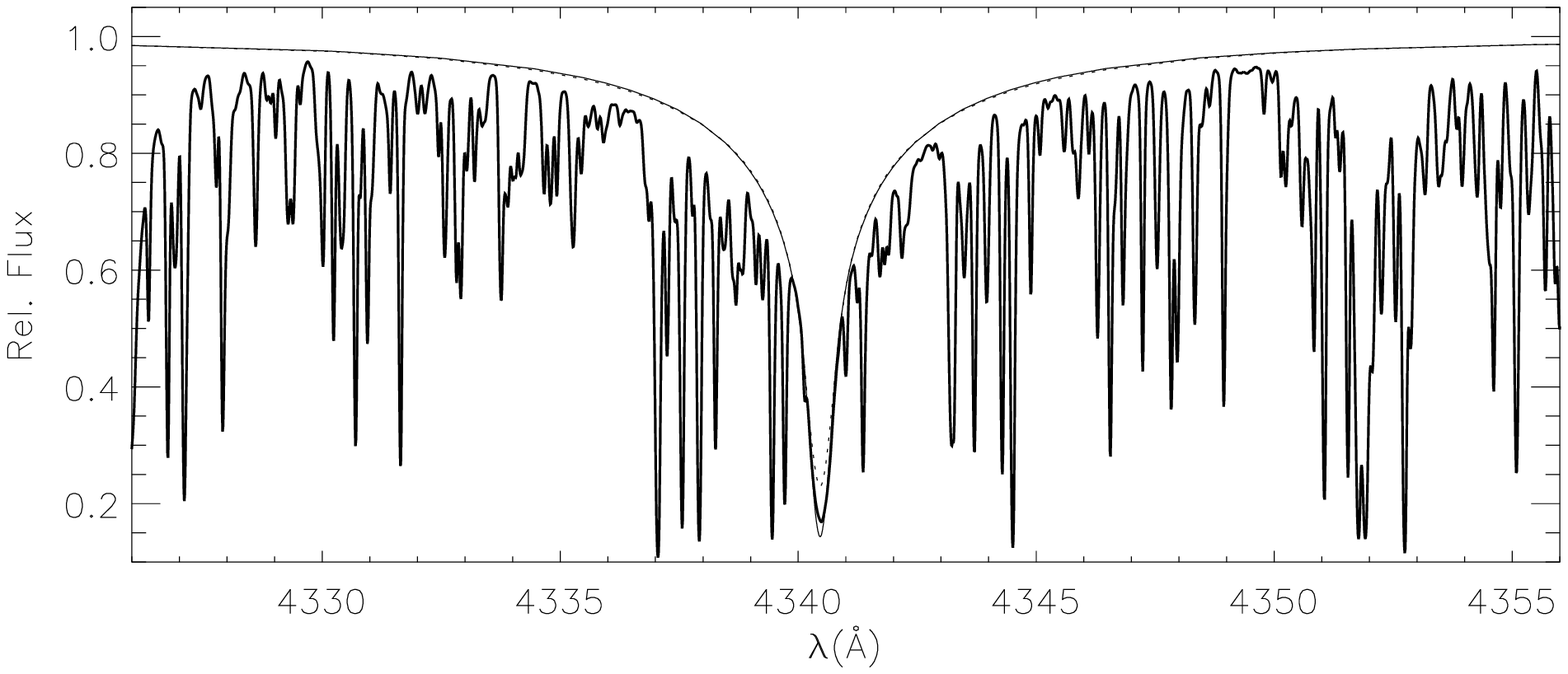}{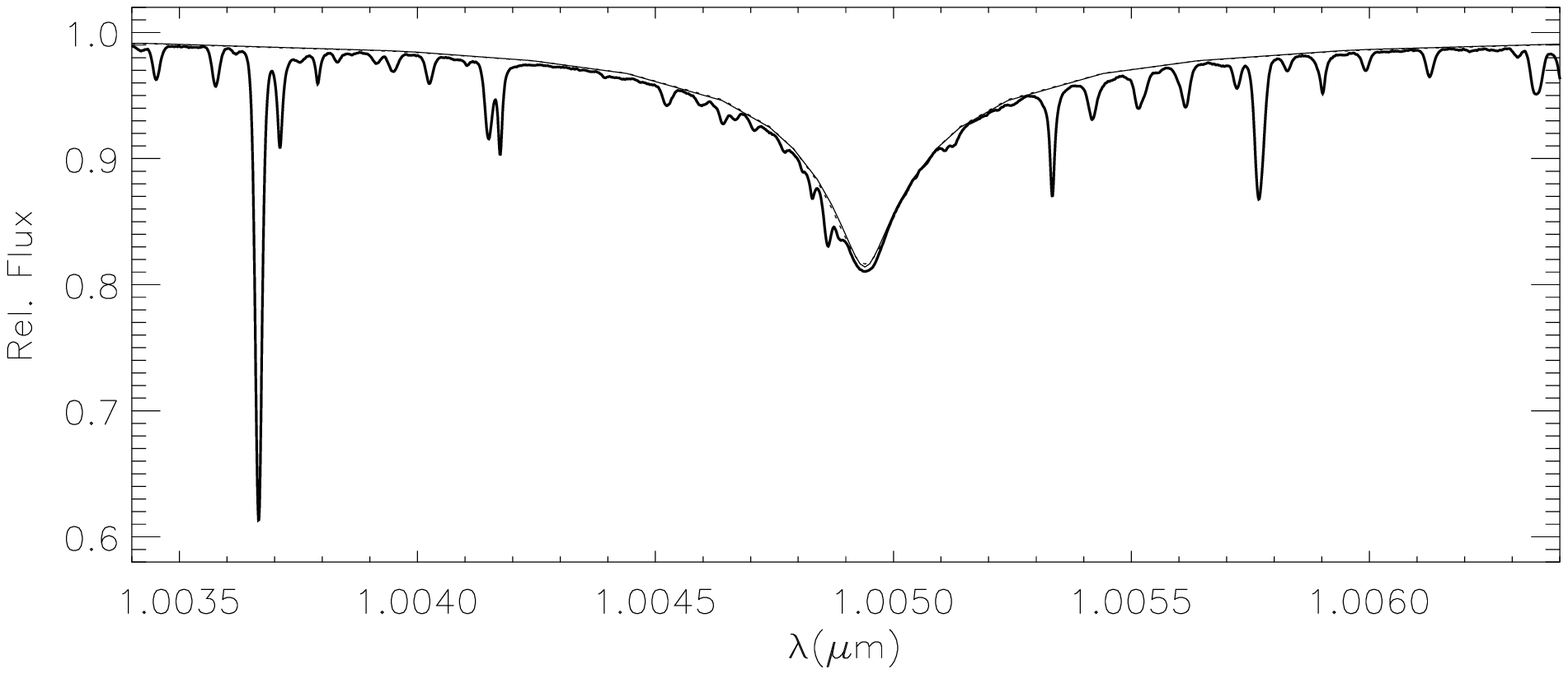}
   \caption{Comparison of non-LTE ({\em full thin line}) and LTE ({\em
dotted line}) synthetic fluxes for the accessible Balmer and Paschen 
series members 
with the normalised NSO/Kitt Peak FTS data from the solar flux atlas 
of Kurucz et al.~(1984, {\em thick line}). Stark and resonance broadening (the
latter for H$\alpha$--H$\gamma$ only) are treated using the tabulations of
SH and BPO. In order to present the best fits, LTE results using the HM
model are shown, and non-LTE predictions using MACKKL. Unphysical LTE
computations using MACKKL give core emission for the Balmer lines, see
Fig.~\ref{hadfig}. In the case of H$\beta$ the MAFAGS-OS model profile, 
neglecting BPO resonance~broadening, is also shown ({\em long-dashed line}).}
    \label{linesfig}%
    \end{figure*}
The non-LTE line-formation computations are carried out using a hybrid
approach. Based on either the classical semi-empirical MACKKL and
HM models or the theoretical SUNK94, A-1D and MAFAGS-OS models
we perform 1D non-LTE computations using {\sc Detail}
and {\sc Surface} (Giddings~1981; Butler \& Giddings~1985).
The coupled radiative transfer and statistical equilibrium equations are solved 
with {\sc Detail}, using a 15-level\,$+$\,continuum version of the recommended model atom for
hydrogen of PB, thus including a more
realistic treatment of electron-collision processes than previously possible.
Inelastic collisions with neutral hydrogen particles are accounted for
using the formula of Steenbock \& Holweger~(1984) with a scaling factor
$S_{\rm H}$\,$=$\,2.
Background opacities are adopted from the {\sc Atlas9} package
(Kurucz~1993), 
except for metal photoionization cross-sections, which are from the Opacity 
Project (Seaton et al.~1992). Line-blocking is included by considering 
Kurucz'(1993) Opacity Distribution Functions. The emergent flux is computed with 
{\sc Surface}, either based on non-LTE populations or in LTE.
Line-broadening by charged and neutral perturbers is accounted for by Stark
profiles from SH and the self-broadening formalism of BPO.

Note that we cannot provide a fully consistent
treatment of the problem with our restricted non-LTE approach, and we will miss
the subtle effects introduced by a full 3D line-formation computation
(e.g.~Asplund et al.~2000). However, 
important conclusions can be drawn despite these restrictions, as will be
shown next.


\section{The solar Balmer and Paschen lines}

\begin{figure}
\plotone{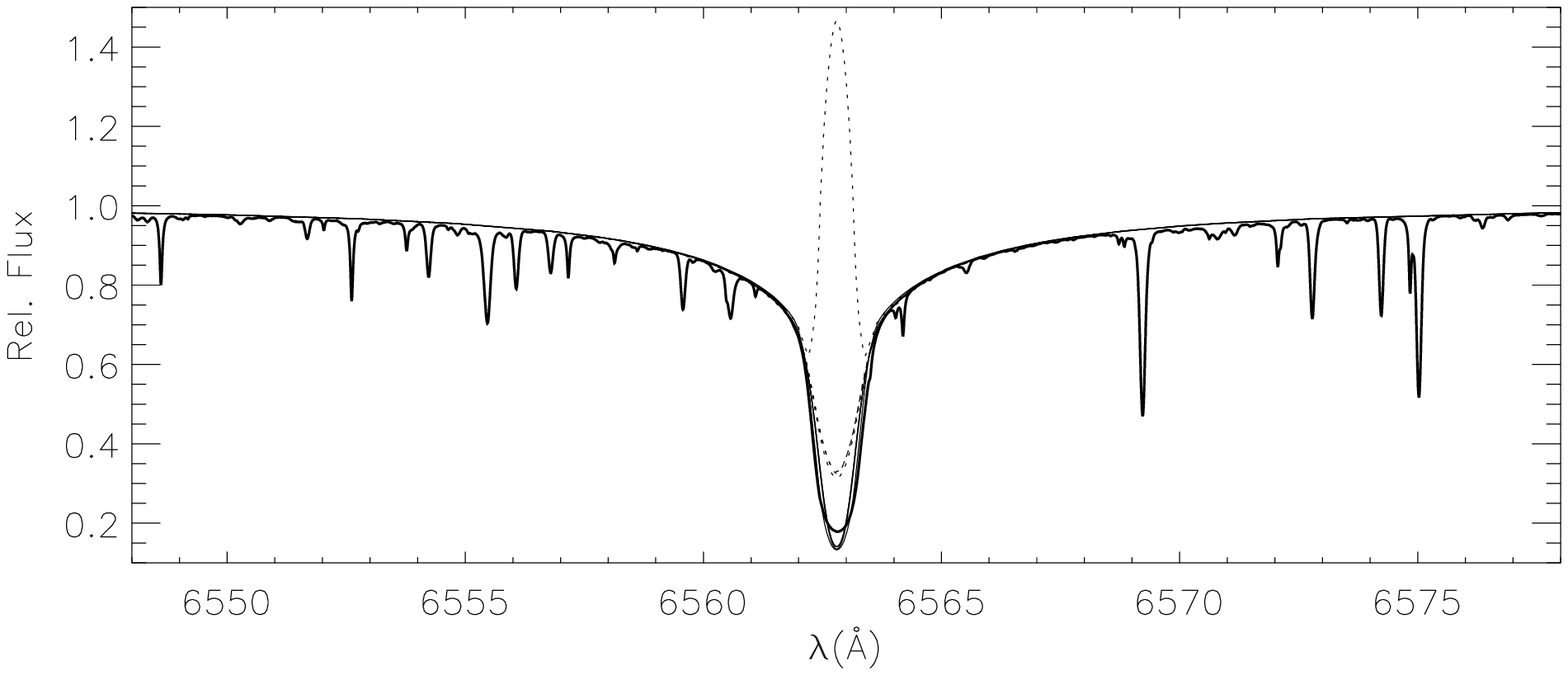}
\plotone{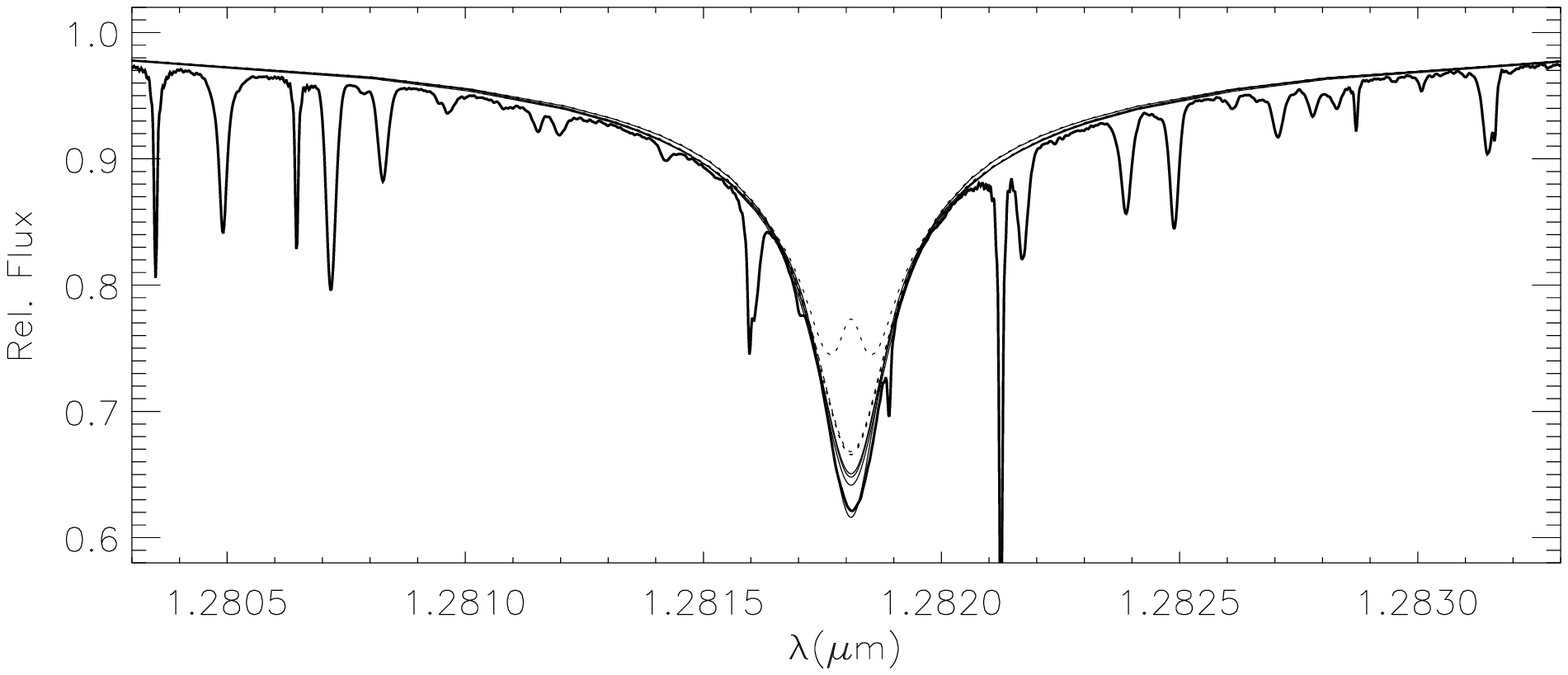}   
   \caption{
   Comparison of four model predictions in non-LTE (full thin lines) and
   LTE (dotted lines) with observation (full thick line). The LTE model with
core emission is MACKKL. Accounting for non-LTE effects eliminates
the emission from the chromosphere. Except for this the
four models predict H$\alpha$ profiles (upper panel) differing only in small
details, while for P$\beta$ (lower panel) the differences are more noticeable.
For the latter MACKKL gives the extremes, resulting in a close to perfect
fit in non-LTE. MAFAGS-OS profiles (not shown) slightly overestimate the 
depths of the wings in P$\beta$ and in H$\alpha$ when accounting for BPO 
resonance broadening.
   }
   \label{hadfig}%
   \end{figure}

The resulting best fits from our non-LTE and LTE computations are compared 
with the normalised NSO/Kitt Peak FTS data from the solar flux atlas
of the quiet solar photosphere by Kurucz et al.~(1984) 
in Fig.~\ref{linesfig}. 
The overall improvement of the profile fits when accounting for
non-LTE effects is evident. Large effects are found for the H$\alpha$ line
core, and a progressive decrease of the non-LTE strengthening towards the higher
Balmer and Paschen lines. 
Only small discrepancies remain, notably at the very line centres
and the well-known problems in the wings of H$\beta$ and the higher Balmer lines.
The former probably result from slight inaccuracies in the solar temperature
structure, requiring a cooler outer photosphere, see also Fig.~1 of Asplund
et al.~(2004), or, alternatively, a modified photosphere-chromosphere
transition as suggested by Avrett~(2003). 
The synthetic non-LTE profiles from the other models vary only slightly from the
best fits, most notably in the very line cores, as shown in
Fig.~\ref{hadfig} exemplarily for H$\alpha$ and P$\beta$, with the exception of MAFAGS-OS which
gives a slightly different solution. As shown for H$\beta$ the line
wings are significantly strengthened and the MAFAGS-OS model has the potential to
resolve the long-standing discrepancy between observed and computed line
wings of the blue Balmer lines. This is due to the higher
continuum flux resulting from the opacity sampling approach in that case.

We thus show that accounting for non-LTE line formation practically removes
the main discrepancies between theory and observation. 
We conclude, that non-LTE analysis of the hydrogen spectrum is mandatory
to recover the physical structure of the solar atmosphere, 
and consequently the basic stellar
parameters. This in turn makes non-LTE Balmer profile fitting 
attractive as a technique to determine the stellar parameters of cool stars in general,
since it is a prototype of this class of stars. It is more powerful
than the Balmer wing fitting technique (Fuhrmann et al.~1993, 1994), as it
probes the atmosphere to a far greater extent. The method even gains in
accuracy, if the Paschen lines can be also accounted for, since they originate
from a level of different excitation energy. However, the problem is
more complicated, as the quantitative interpretation of our
findings will show.

\begin{figure}
   \plotone{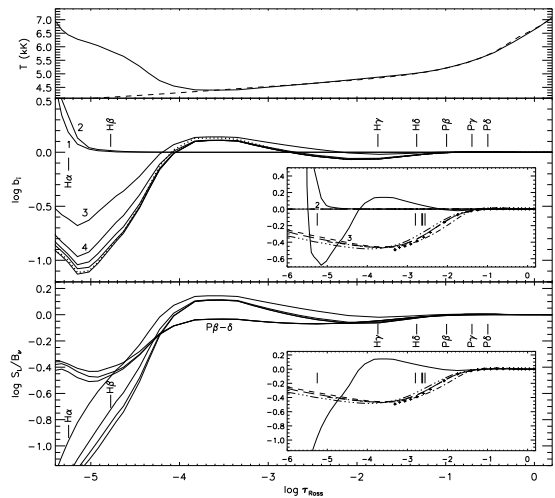}
   \caption{Upper panel: Temperature structure for the MACKKL solar atmospheric model
({\em full line}) and for SUNK94 ({\em dashed}).
Middle panel: Run of hydrogen departure coefficients $b_i$ as a function of
Rosseland optical depth in the MACKKL model. The curves are labelled with the 
level's principal quantum number $n$. Energetically higher levels converge to the
continuum behaviour ({\em dotted line}). The depths for line-core formation are
indicated. The insert displays departure coefficients for $n$\,$=$\,2,3 for 
all the model atmospheres
considered (A-1D: {\em dash-dotted}; MAFAGS-OS:
{\em long-dashed}; HM: {\em triple-dot-dashed}; MACKKL
without chromosphere: {\em crosses}).
Line-core formation depths for H$\alpha$ are marked, the H$\alpha$ core 
in MACKKL is formed in the chromosphere while in the other
stratifications this is shifted to the deeper photosphere. 
Lower panel: Ratio of line source function $S_{\rm L}$ to Planck function 
$B_{\nu}$ at line centre. Non-LTE
strengthening of the Balmer and Paschen line cores occurs due to the relative 
underpopulation of the transitions' upper levels. The graphs are
qualitatively similar for the five cases without chromosphere, while some
differences occur in the details, see insert (for H$\alpha$ only). 
The MACKKL model gives a
qualitatively different idea of the line-formation process, while the
resulting profiles are almost identical, see Fig.~\ref{hadfig}.
}
   \label{theofig}%
   \end{figure}

We begin by discussing the departure coefficients $b_i$\,$=$\,$n_i/n_i^{\ast}$ (the $n_i$ and
$n_i^{\ast}$ being the non-LTE and LTE populations of level $i$, respectively).
These are displayed in the inset of Fig.~\ref{theofig}, 
for the atmospheres without chromosphere. 
The ground state and the first excited level stay in detailed balance
throughout almost the
entire Balmer line-forming depths in the solar atmosphere, due to the optical
thickness of the Ly$\alpha$ transition at those depths. 
The $n = 3$ and higher levels are in
detailed balance deep in the photosphere, but develop 
a non-LTE underpopulation further out.
However, the levels with higher $n$-values stay
in detailed balance {\em relative} to each other at these atmospheric
depths, and they also collisionally couple tightly to the continuum.
Inspection of the non-LTE line core source functions $S_{\rm L}$ in the
inset of Fig.~\ref{theofig} indicates a drop of $S_{\rm L}$ below the Planckian value and thus non-LTE
strengthening of the lines.

The marked reduction of the line centre intensities
in these cases is caused exclusively by
photon escape (see e.g. Mihalas~1978, Ch.~11-2). In fact, it is photon
escape from H$\alpha$ itself which controls the non-LTE departures of hydrogen: when
setting H$\alpha$ into detailed balance, the non-LTE effects on {\em all}
levels of hydrogen vanish, and the LTE line profiles are recovered in that
case.   
The importance of this radiative transition for the non-LTE problem 
is due to the failure of collision processes to establish detailed
equilibrium between the $n$\,$=$2 and 3 levels, as the energy gap of 1.89\,eV
implies that only particles in the high-velocity tail of the Maxwell
distribution will be relevant. For all other transitions between the
energetically higher levels the gaps are 0.66\,eV, or lower, thus they are
easily coupled via collisions at the temperatures prevailing in the solar
atmosphere.

The behaviour of the departure coefficients and line source functions for all 
the solar model stratifications without chromosphere is qualitatively similar, see the insets in Fig.~\ref{theofig}, 
with only minor differences occur in the details.

The MACKKL model on the other hand shows a fundamentally different 
behaviour, our results agreeing closely with those of Avrett (priv.~comm.,~2004).
Here an overpopulation of the $n=3$ level occurs around the solar 
temperature minimum and a marked
underpopulation in the chromosphere, where the cores of the H$\alpha$ and
H$\beta$ are formed as seen in Fig.~\ref{theofig}.
We attribute this to the irradiation of the photosphere by energetic photons
from the high-temperature plasma of the chromosphere, a classical non-LTE
situation which is accounted for in the construction of the MACKKL model (see Vernazza,
Avrett \& Loeser~1981). 
In order to test this, we
artifically restrict the MACKKL model to depths below the temperature
minimum, thus reproducing the non-LTE behaviour of the models
without chromosphere in the same qualitative way. H$\alpha$ line-formation then 
takes place exclusively in the photosphere.

However, the profiles with or without chromosphere are in all cases almost 
identical and show equally good agreement with the observed line profiles
(core emission is characteristic for (unphysical) LTE calculations using
MACKKL). 

The differences in the model predictions produce more noticeable effects in
P$\beta$, which probes the deeper solar photosphere. Distinguishing the
different models is possible for the high quality spectrum available for the
Sun -- for stellar analyses, where compromises with regard to resolution and
S/N have to be made in most instances, such efforts may be difficult. 


\section{Implications for quantitative spectroscopy of cool stars}
   \begin{figure}
   \plotone{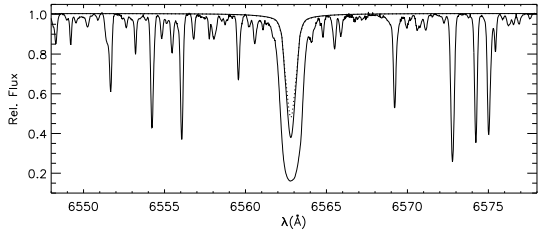}
   \caption{Synthesis of H$\alpha$ for the mildly metal poor red
giant Arcturus, using a conventional LTE model atmosphere without chromosphere
(Peterson et al.~1993) with non-LTE ({\em thin full line}) and LTE ({\em
dotted}) line formation: pointing towards an inadequacy of such models for
quantitative spectroscopy.
%
%
}
   \label{arcturus}%
   \end{figure}
Chromospheres are an integral part of the atmospheres of cool stars.
Their impact on the hydrogen line-formation and atmospheric structure will
gain in importance where chromospheres are more pronounced, 
e.g. young stars, or more generally, objects showing chromospheric activity, and
in situations which
facilitate departures from LTE, i.e. for metal-poor or evolved objects.
The mechanisms will be the same as for the Sun: irradiation by 
energetic photons from the chromospheres modifies the statistical
equilibrium of hydrogen throughout the stellar photosphere, however leading
to more pronounced deviations from LTE. More important than
the changes of the level populations, which determine the line strengths, is
the effect on the ionization balance and thus the pool of free electrons.
This in turn controls the formation of H$^-$, the main opacity source in 
cool stars, and consequently the stellar continuum. Metal-poor red giants
will therefore show pronounced effects, as the role of hydrogen as electron
source is strengthened with regard to the metals.

The primary target to verify this is the mildly metal-poor K-giant
\objectname{Arcturus}, which can be studied in great detail due to its proximity, 
and allows for a model-independent determination of basic stellar
parameters (Griffin \& Lynas-Gray~1999). Synthetic H$\alpha$ profiles from
LTE and non-LTE line-formation calculations using a conventional model
atmosphere without chromosphere (Peterson, Dalle Ore \& Kurucz~1993)
are compared to observation (Bagnulo et al.~2003) in Fig.~\ref{arcturus}.
The failure to obtain even a rudimentary fit indicates an inadequacy of
this type of model atmosphere for the quantitative spectroscopy of such stars.
Note that the discrepancy in the surface gravity of the Peterson et al.
model with that derived by Griffin \& Lynas-Gray has no impact on the problem,
as the hydrogen lines are insensitive to variations of
$\log g$ in this region of parameter space.  
Unfortunately, a chromospheric model by Ayres \&
Linsky~(1975) does not include the deeper photosphere, so that a
quantitative investigation is beyond the scope of the present work.
Without doubt, a crucial ingredient, and a boundary condition, for successfully deriving
realistic model atmospheres will be Balmer profile fits.

Quantification of the systematic errors introduced by the use of
conventional model atmospheres in present-day cool star analyses is
a necessity, but will require the construction of semi-empirical non-LTE model
atmospheres with chromospheres for stars other than the Sun in a first step
and ultimately a theoretical understanding of chromospheres
from first principles. If shown to be non-negligible, this will have
far-reaching consequences since 
the properties of these stars (fundamental parameters, abundances) are the principal sources 
of our knowledge of galactic evolution (including the solar neighbourhood, globular
clusters, galactic bulges, dwarf galaxies etc., to the properties of giant
ellipticals from their integrated light) and cosmology (the first step in the 
calibration of the galactic distance scale, which relies predominantly on
late-type stars; visible baryonic matter content through modified $M/L$-ratios).


%

\acknowledgments
We would like to thank C. Allende Prieto for pointing us to the solar
Balmer line-core discrepancy. We would also like to thank T.~Gehren for many
stimulating discussions on cool stars, and our referee, E.H.~Avrett, for
helping us to improve on the manuscript.
The NSO/Kitt Peak FTS data used here were produced by NSF/NOAO.

\clearpage

\end{document}